# Forecasting Nigerian Equity Stock Returns Using Long Short-Term Memory Technique

## Adebola K. Ojo [a*] and Ifechukwude Jude Okafor [a]

*[a] Department of Computer Science, University of Ibadan, Ibadan, Nigeria.*

***Authors' contributions***

*This work was carried out in collaboration between both authors. Both authors read and approved the final manuscript.*

*Original Research Article*

## Abstract

Investors and stock market analysts face major challenges in predicting stock returns and making wise investment decisions. The predictability of equity stock returns can boost investor confidence, but it remains a difficult task. To address this issue, a study was conducted using a Long Short-term Memory (LSTM) model to predict future stock market movements. The study used a historical dataset from the Nigerian Stock Exchange (NSE), which was cleaned and normalized to design the LSTM model. The model was evaluated using performance metrics and compared with other deep learning models like Artificial and Convolutional Neural Networks (CNN). The experimental results showed that the LSTM model can predict future stock market prices and returns with over 90% accuracy when trained with a reliable dataset. The study concludes that LSTM models can be useful in predicting financial time-series-related problems if well-trained. Future studies should explore combining LSTM models with other deep learning techniques like CNN to create hybrid models that mitigate the risks associated with relying on a single model for future equity stock predictions.

_________________________________________

*\*Corresponding author: Email: adebola_ojo@yahoo.co.uk, adebolak.ojo@gmail.com, ak.ojo@ui.edu.ng;*

*Keywords: Long-short term memory; deep neural networks; forecasting; convolutional neural networks; performance evaluations.*

# 1 Introduction

Despite the inherent profits expected investments in the stock market are associated with considerable financial risks. This is owing to the stock market's tumultuous non-linearity and unpredictable nature, as well as other fundamental and technical elements that influence it. The innate difficulty of properly and sufficiently determining future stock market returns has drawn the attention of numerous academic and corporate institutions over the years to provide ways for avoiding the hazards outlined [1].

Many researchers have in the recent past used traditional statistical and mathematical methodologies to forecast stock price fluctuations, and considerable successes were made, but not outstanding enough. Recently, there has been a significant progress recorded with the introduction of the Machine Learning paradigm in dealing with the non-linearity characteristic of financial time series such as stock market predictions. Many artificial intelligence and machine learning research papers are focusing on the inherent potentials of deep Neural methods for forecasting financial time series forecasting challenges like stock market predictions.

Developed in 1997 by Hochreiter and Schmid Huber, LSTM is a part of the Recurrent Neural network family algorithm and amongst many of its usefulness, can be used to predict non-linear financial time-series trends of a stock movement or returns [2].

With the intent of contributing to the body of knowledge, this study aims at investigating the potential of a Neural Network (Deep learning) technique known as LSTM to address the lingering challenges of predicting stock market returns (via stock price movements or trends) and propose a model to forecast viable future stock price trends that would yield great returns to stock market investors.

# 2 Literature Review

Several works have been done to proffer reliable solutions to challenges facing the predictability of equity stock movements to enable a typical investor to make informed decisions. These informed decisions in turn give the investor the leverage to create wealth. Some works or studies done by scholars ranging from the application of conventional statistical and analytical tools to the modern-day application of Artificial intelligence (machine learning and advanced neural networks) models and techniques were highlighted as follows:

In [3], the equity market's significance in the investment world cannot be over-emphasized. The increment or decrement shift in stock price trends can influence the decision-making disposition of a typical investor. In the early years of stock market forecasting, predicting models deployed are seemingly linear. However, recent studies have shown an increase in the use of deep learning techniques for the prediction of stock market returns. In this study, two deep learning architectures (RNN and LSTM) were deployed as models for the forecasting of Indian national stock exchange listed companies. The result showed commendable performance of the deep learning hybrid models deployed. The long short-term memory model performance accuracy was outstanding.

Adebiyi et al. [4] opined that Stock market prediction is mostly characterized using technical indices alone, however, in their study, they chose to deviate away from the conventional norm by applying machine learning methods (Top-level random forest (TRF), logistic regression (LR), Support vector machine (SVM) and Neural networks (NN)) on equity stock trading. The experimental study outcome shows TRF prediction was more accurate in performance than other algorithms tested.

The Stock market's complex nature has made its predictability a daunting task. The above obvious challenges posed have made conventional batch processing prediction method ineffective [5]. In this study, their proposed online machine learning technique used both Recurrent Neutral Network (RNN) and LSTM architectures on Indian Market stock data. The proposed online network was trained and evaluated for accuracy and results tabulated. The dual deep neural networks models built performed better in accuracy (in comparison to earlier conventional batch-processing models for predicting stock price outlooks.

Stock market investment is of great economic importance to individuals and the world at large. However, its volatile uncertainties need efficient forecasting tools. Godknows and Olusanya [6] carried out a study on stock price prediction using RNN by collecting Google stock price dataset of over ten years. The model built indicated RNN can predict the future stock price returns correctly in short term basis.

Ibidapo et al. [7] showed financial time series related problems, forecasting future equity stock values and returns needs robust mathematical or technological models and tools. This is due to the non-linear nature of financial time series challenges and its attendant volatility and uncertainties. The challenges posed by the complexity of equity stocks have drawn the attention (interest) of researchers, investors, and academia community. Ibidapo et al. [7] carried out a study on Brazilian national stock prices by evaluating the performance predictability of LSTM neural network. The results show LSTM predicted accurately the future stock movement returns of selected equity stocks under test.

Equity Share Market investment entails monitoring and managing investment risks, and not attempting to evade it. With the above mindset [8], a passionate knowledgeable investor would be wise enough to seek effective predicting tools with high accuracy that can help with informed decision for analyses and deduction on what stock to buy or sell or to hold until later time in the future. Chanddrika and Sreenvasan [8] combined technical analysis and machine learning architecture to predict the performance of the hybrid tools used on stocks datasets. The result showed the hybrid models proposed did well in the forecast exercise.

Having evaluated the non-linearity nature of financial market, (both commodities and stock market) [9] proposed the use of RNN model to tackle the challenge of predicting accurately (via relevant datasets) the future values of stock market. The RNN model built was trained, validated, and tested using NSE stock prices, result revealed via the graphs plotted and obtained that the recurrent neural network variant predicted accurately the future price returns of selected equities.

**Long Short-Term Memory (LSTM) Architecture:**

LSTM is a variant form of RNN that was invented in 1997 to help solve (and correct) the limitations of a typical recurrent neural network called Vanilla recurrent neural network. A typical LSTM structure is depicted in Fig. 1.

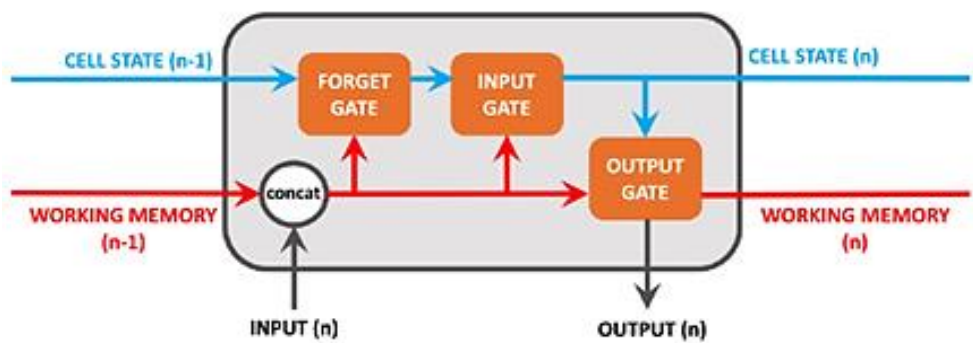

**Fig. 1. A sample of LSTM (RNN variant) [8]**

LSTM uses the cell state at time-interval to predict the next internal value instances. The cell state vector represents the LSTM memory. Usually, the cell state can be altered by specific gates (Forget, Input, or Output gate respectively). A gate is architecturally a sigmoid neural net layer characterized by a point-wise multiplication operator. It coordinates the information direction to/from the LSTM memory. Gates are controlled by concatenating the output from the previous time step and the current input (and optionally the cell state vector.). Forget Gate handles what information (or values) to remove from the working memory. Input Gate determines what new information (value item) will be included in a cell state from the current input. Output Gate conditionally decides what to output (or display) from memory. To update and refresh the LSTM memory, the cell state vector aggregates the two components which are the "old memory" through the forget gate" and the "new memory" through the input gate [10].

Further works on stock market prediction and performance are found in [11] [12], and [13]

## 3 The Methodology

Fig. 2 depicts the architectural design of the proposed LSTM model. It helps laid down step by step the processes (methodologies) of how to go about achieving the objectives of this study as enunciated below:

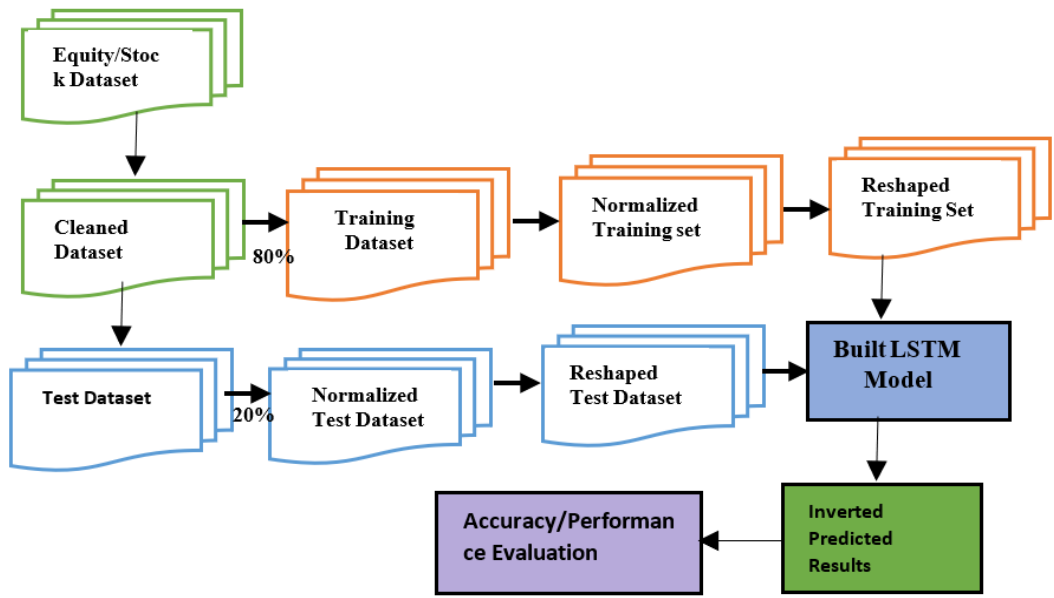

**Fig. 2. Detailed architectural design of the lstm prediction model**

Dataset was obtained from Nigerian Stock Exchange website which is made up of equity-stocks that represent each Nigerian economic sector, such as manufacturing, transportation, and telecommunications as shown in Table 1. Only equities with over 20 years of consistent historical dataset (2001-2022) were chosen for the tests in this study (which include GTBank, AIICO, Zenith Bank, UACN, Flourmill and Wapco etc.).

The dataset was cleaned up by eliminating noise, missing data, and inconsistencies. The features in the dataset consist of Date of transaction, Stock-Name, Opening price, Highest-Intra-day price, Lowest Intra-day, and daily volumes of equity-stock traded for the day. As part of building and implementation of Neural Network model (i.e., LSTM), the dataset scaling and normalization is a critical prerequisite. This entails producing a scalar object (with feature range of 0, 1 values) using the MinMaxScaler tool. Before transforming the dataset to matrix format, dataset reshaping (converting the dataset to a 3D array) is vital for the LSTM algorithm to read and train the proposed model effectively.

The dataset was later split into two matrix variables, namely, "the training and testing" dataset (i.e., Xtrain, and Xtest) respectively. To build the proposed model, the LSTM (Deep learning) algorithm was applied with a unit value of 40, a total input shape of (55,1), and a Dropout value of 0.2.

The model was trained using an LSTM with a batch size of 35 and with an epoch value of 50. 80% of the dataset was used to train the LSTM model. To verify the accuracy of how well-trained the model was, the trained model was test-proved with 20% of the dataset reserved for validation and subsequently appended to the trained dataset. Graphs and charts were plotted with real stock prices and the predicted stock price values against time.

**Table 1. Classification of stocks based on their Nigerian economy sector**

| Sector | Company 1 | Company 2 |
|---|---|---|
| Banking | GTB | Zenith |
| Consumer Goods | Nestle | Flourmill |
| Insurance | AIICO | Cornerstone |
| Oil and Gas | Total | OANDO |
| Industrial Goods | Dangote Cement | WAPCO |
| Breweries | Guinness | NB |
| Agriculture | Okomuoil | Presco |
| Conglomerates | UACN | Transcorp |
| Telecommunications Building | AIRTEL | MTNN |
| Materials | BERGER | CAP |
| Hotels | Ikeja Hotel | Transcorp Hotel |
| Support & Logistics | Caverton | CILEASING |
| Transport | NAHCO | SKYAVN |

**Table 2. A sample of GTB equity stock movement**

| A | B | C | D | E | F | G |
|---|---|---|---|---|---|---|
| date | symbol | open | close | low | high | volume |
| 02/01/2001 | GTB | 4.17 | 4.17 | 4.1 | 4.17 | 1321000 |
| 03/01/2001 | GTB | 4.37 | 4.35 | 4 | 4.37 | 2491041 |
| 04/01/2001 | GTB | 4.56 | 4.56 | 4.4 | 4.56 | 2579950 |
| 05/01/2001 | GTB | 4.7 | 4.35 | 4.34 | 4.7 | 1118861 |
| 08/01/2001 | GTB | 4.4 | 4.25 | 4.25 | 4.5 | 367466 |
| 09/01/2001 | GTB | 4.45 | 4.21 | 4.2 | 4.45 | 1089925 |
| 10/01/2001 | GTB | 4.3 | 4.4 | 4.21 | 4.42 | 1468750 |
| 11/01/2001 | GTB | 4.52 | 4.57 | 4.5 | 4.59 | 1529660 |
| 12/01/2001 | GTB | 4.79 | 4.78 | 4.58 | 4.79 | 2615062 |
| 15/01/2001 | GTB | 5.01 | 5.01 | 5.01 | 5.01 | 942804 |
| 16/01/2001 | GTB | 5.26 | 5.12 | 4.86 | 5.26 | 4310481 |
| 17/01/2001 | GTB | 5.37 | 5.37 | 5.21 | 5.37 | 3951523 |
| 18/01/2001 | GTB | 5.63 | 5.38 | 5.37 | 5.63 | 1421149 |
| 19/01/2001 | GTB | 5.12 | 5.12 | 5.12 | 5.12 | 663687 |
| 22/01/2001 | GTB | 4.87 | 4.87 | 4.87 | 4.87 | 246962 |
| 23/01/2001 | GTB | 4.63 | 4.8 | 4.63 | 4.8 | 436621 |
| 24/01/2001 | GTB | 4.81 | 4.56 | 4.56 | 4.87 | 5228938 |
| 25/01/2001 | GTB | 4.56 | 4.72 | 4.4 | 4.72 | 1092999 |

**Developmental Tools Utilized for the LSTM Model Implementation:**

The developmental tools used to implement the LSTM model are Python 3.11, Anaconda, Jupyter Notebook. Other Python imported libraries are: Keras, Tensorflow, Pandas, Numpy, Scikit-learn and Matplotlib libraries for data and graphic visualizations.

**Evaluation of the LSTM Model:**

In the evaluation of the LSTM model built, three methods were used:

i.) The developed LSTM model was subjected to performance evaluation using evaluation metrics.
ii.) The LSTM model was benchmarked with other developed deep neural models.
iii.) The result of this study was benchmarked with the results of other research studies.

**Evaluations Metrics Utilised:**

The evaluation metrics used were Accuracy, Mean Absolute Error (MAE), Mean Squared Error (MAE), and Root Mean Squared Error.

## 4 Results and Discussion

The dataset used for training the LSTM model developed was split into two (training and testing), of which 80% was used for training the LSTM model and 20% was used for testing (predicting and evaluating) the LSTM model built. After the feeding the LSTM model built with the necessary parameters (e.g. Optimizer, learning rate, batch size, no of iteration or epoch), the next step was to train the LSTM Model with the "training" dataset and then, allow the model on its own to do the prediction.

```
In [43]: model_1.summary()
Model: "sequential_2"
_________________________________________________________________
Layer (type)                 Output Shape              Param #
=================================================================
lstm (LSTM)                  (None, 30, 70)            20160
lstm_1 (LSTM)                (None, 30, 70)            39480
lstm_2 (LSTM)                (None, 70)                39480
dense_7 (Dense)              (None, 1)                 71
=================================================================
Total params: 99,191
Trainable params: 99,191
Non-trainable params: 0
_________________________________________________________________
```

**Fig. 3. Summarized features of the LSTM model**

|   | date | symbol | open | close | low | high | volume |
|---|------|--------|------|-------|-----|------|--------|
| 0 | 04/05/2000 | GTB | 2.26 | 2.30 | 2.266 | 2.30 | 200200 |
| 1 | 05/05/2000 | GTB | 2.35 | 2.35 | 2.270 | 2.35 | 636701 |
| 2 | 08/05/2000 | GTB | 2.38 | 2.35 | 2.350 | 2.38 | 567700 |
| 3 | 09/05/2000 | GTB | 2.40 | 2.46 | 2.350 | 2.46 | 268500 |
| 4 | 10/05/2000 | GTB | 2.46 | 2.34 | 2.340 | 2.46 | 971000 |

**Fig. 4. Dataset of GTBank**

|  | open | close | low | high | volume |
|---|------|-------|-----|------|--------|
| count | 5081.000000 | 5081.000000 | 5081.000000 | 5081.000000 | 5.081000e+03 |
| mean | 20.511019 | 20.512726 | 20.272076 | 20.747585 | 1.442003e+07 |
| std | 10.446109 | 10.457982 | 10.340171 | 10.553160 | 1.991220e+07 |
| min | 1.690000 | 1.690000 | 2.050000 | 1.690000 | -4.210300e+04 |
| 25% | 12.500000 | 12.500000 | 12.300000 | 12.630000 | 3.323861e+06 |
| 50% | 19.330000 | 19.290000 | 19.050000 | 19.650000 | 9.097378e+06 |
| 75% | 28.010000 | 28.000000 | 27.950000 | 28.400000 | 1.857721e+07 |
| max | 54.710000 | 54.710000 | 53.000000 | 57.000000 | 4.948477e+08 |

**Fig. 5. Shows the statistical analysis of GTBank Dataset**

Fig. 5 shows the dataset of GTBank as well as its statistical analysis in terms of mean, maximum and minimum values in the dataset.

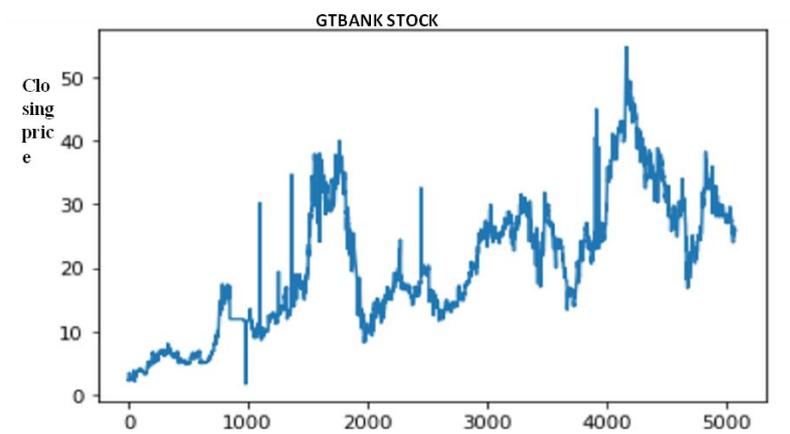

**Fig. 6. Plot of GTBank's dataset closing prices (y-axis) plotted against time (x-axis)**

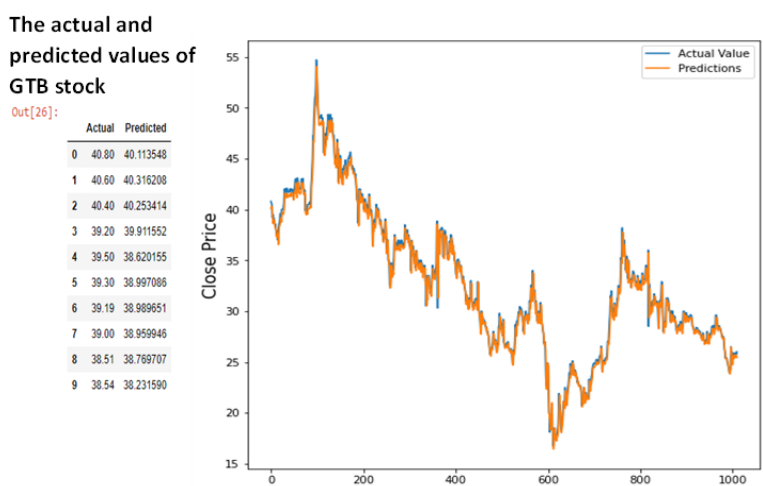

**Fig. 7. GTBank's actual closing prices plot against the predicted prices**

**Table 3. Predictions of the LSTM model on few selected equity-stocks**

| Stock Name | Accuracy | MAE | MSE | RMSE |
|---|---|---|---|---|
| GTBank | 0.98533 | 0.5672 | 0.7624 | 0.87318 |
| CAP | 0.967 | 0.9025 | 1.5352 | 1.2590 |
| Flourmill | 0.9777 | 0.5192 | 0.9283 | 0.9634 |
| BERGER | 0.9394 | 0.1365 | 0.0770 | 0.2776 |
| AIICO | 0.9346 | 0.0373 | 0.00285 | 0.544 |
| UACN | 0.9788 | 0.2490 | 0.2960 | 0.5442 |
| WAPCO | 0.9739 | 4.474 | 2.115 | 0.7401 |
| ZENITH | 0.96730 | 0.3908 | 0.3860 | 0.6321 |

**Interpretation of the LSTM model performance using standard regression metrics:**

To measure the performance of the LSTM model, regression-related evaluation metrics such as accuracy, Mean Absolute Error (MAE), Mean Square Error, Root Mean Squared Error (RMSE) were used.

**Table 4. Comparison between the LSTM model against other Neural Network models**

| Nigerian Stocks | Accuracy | Deep Neural Network Models | | | | | | | | | | |
|---|---|---|---|---|---|---|---|---|---|---|---|---|
| | | Long Short Term Memory | | | Artificial Neural Network (ANN) | | | | Convolutional Neural Network (CNN | | | |
| | | MAE | MSE | RMSE | ACCURACY | MAE | MSE | RMSE | ACCURACY | MAE | MSE | RMSE |
| GTBAN | 0.98533 | 0.5672 | 0.7624 | 0.87318 | 0.9848 | 0.592 | 0.7984 | 0.8935 | 0.9783 | 0.5372 | 0.7524 | 0.85318 |
| CAP | 0.967 | 0.9025 | 1.5352 | 1.259 | 0.9768 | 0.6242 | 1.1117 | 1.058 | 0.965 | 0.9225 | 1.5252 | 1.249 |
| FLOURMILL | 0.9777 | 0.5192 | 0.9283 | 0.9634 | 0.9732 | 0.6183 | 1.1197 | 1.058 | 0.9677 | 0.5092 | 0.9283 | 0.9634 |
| BERGER | 0.9394 | 0.1365 | 0.077 | 0.2710 | 0.9412 | 0.1022 | 0.074 | 0.2732 | 0.9384 | 0.1355 | 0.077 | 0.2776 |
| AIICO | 0.9346 | 0.0373 | 0.00285 | 0.544 | 0.9554 | 0.026 | 0.0194 | 0.044 | 0.9336 | 0.0373 | 0.00285 | 0.544 |
| UACN | 0.9788 | 0.249 | 0.296 | 0.5442 | 0.9726 | 0.3532 | 0.3834 | 0.6192 | 0.9778 | 0.249 | 0.296 | 0.5442 |
| VAPC | 0.97739 | 4.474 | 2.115 | 0.7401 | 0.9701 | 0.69 | 5.06 | 2249 | 0.9719 | 4.471 | 2.115 | 0.7301 |
| ZENITH | 0.9673 | 0.3908 | 0.386 | 0.6321 | 0.9719 | 0.3217 | 0.3317 | 0.576 | 0.9653 | 0.3918 | 0.376 | 0.6231 |

Using the standard evaluation metrics mentioned above, the first row of Table 2 can be interpreted thus, The LSTM model predicted GTBank equity stock returns with high accuracy of 0.98 (98%), while MAE of value 0.567, MSE value of 0.762 and RMSE of 0.87 respectively shows the loss errors are low, thus a good prediction performance by the LSTM model. Other equity stocks were well predicted too by the model.

**Benchmarking the LSTM model against other Neural Network models:**

As part of the objectives of this study the LSTM model was benchmarked with other Deep Neural Networks. Table 4 showed that LSTM performance was impressive in prediction process.

In addition, Fig. 8 depicts the graphical comparison of the LSTM model's performance accuracy against other Neural Network models in the prediction of the few selected equity-stocks.

The graph shows pictorially that the LSTM model built did extremely well against some other neural network models such convolutional neural network (CNN), and Artificial Neural Networks (ANN) respectively.

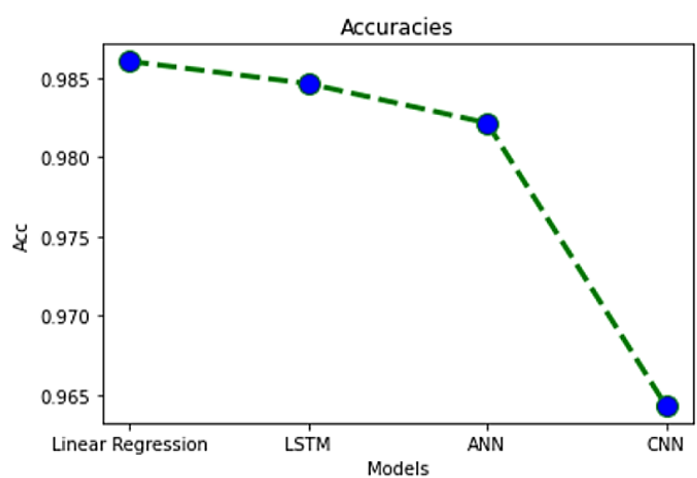

**Fig. 8. Graphical comparison of the LSTM model with other deep neural networks**

## 5 Conclusion

The deep long-short memory model developed in this study performed credibly well in predicting most Nigerian equity stocks when subjected to reliable historical training dataset. This LSTM model has an average of 97% predictive accuracy and 3% lost error so far in the experiment and findings. It equally measured well when bench-marked with other advanced deep learning neural networks such as Convolutional and Artificial Neural networks. The findings show that LSTM models only need well-cleaned and normalized huge dataset to perform at its best.

In future works, there is a need to combine two or more deep learning neural networks together to form hybrids to forestall or mitigate the high risk of relying on a single model to predict equity-stocks returns accurately. It would therefore be technologically benefiting in experiencing technological advancement when more hybrids of deep neural models are developed to harness the potential advantageous features and characteristics inherent in hybrid technological solutions.

## Disclaimer (Artificial Intelligence)

Author(s) hereby declare that NO generative AI technologies such as Large Language Models (ChatGPT, COPILOT, etc) and text-to-image generators have been used during writing or editing of manuscripts.

## Competing Interests

Authors have declared that no competing interests exist.